\documentclass[prl,twocolumn,showpacs,amsfonts,floatfix]{revtex4}
\usepackage[dvips]{graphicx}
\usepackage{bm}
\newcommand{\Journal}[4]{#1 {\bf #2}, #3 (#4)}
\newcommand{\PR}{Phys. Rev.}
\newcommand{\PRL}{Phys. Rev. Lett.}
\newcommand{\PRA}{Phys. Rev. A}
\newcommand{\JMP}{J. Math. Phys.}
\newcommand{\EPJD}{Eur. Phys. J. D}

\newcommand{\Science}{Science}

\begin{document}
\title {Dynamics of Lieb-Liniger gases}
\author{M. D. Girardeau}
\email{girardeau@optics.arizona.edu}
\affiliation{Optical
Sciences Center, University of Arizona,
Tucson, AZ 85721}
\date{\today}
\begin{abstract}
It is proved that the Lieb-Liniger (LL) cusp condition implementing
the delta function interaction in one-dimensional Bose gases is dynamically
conserved under phase imprinting by pulses of arbitrary spatial form 
and the subsequent \emph{many-body} 
dynamics in the thermodynamic limit is expressed approximately in terms of 
solutions of the time-dependent \emph{single-particle}
Schr\"{o}dinger equation for a set of time-dependent orbitals evolving
from an initial LL-Fermi sea. As an illustrative 
application, generation of gray solitons in a LL gas on a ring by a phase 
imprinting pulse is studied. 
\end{abstract}
\pacs{03.75.-b,05.30.Jp}
\maketitle
Recent advances and experiments on ultracold atomic vapors in atom waveguides 
\cite{Key,Muller,Dekker,Thy,Hinds,Bongs,Hansel,Denschlag,Schreck,Greiner,Arlt,DePMcCWin99,GorVogLea01} 
and potential applicability to atom interferometry \cite{Berman} 
and integrated atom optics \cite{Dekker,Sch} 
create a need for accurate theoretical modelling of such systems, and this
is underscored by recent experimental achievement of optical atom
interferometric structures of Mach-Zehnder and Michelson type \cite{Dumke}.
At sufficiently low temperatures and densities and high transverse frequencies
$\omega_0$ the transverse degrees of
freedom of a Bose-Einstein condensate (BEC) in an atom waveguide
are frozen in the ground transverse mode and the dynamics is described 
by an effective 1D Hamiltonian
with delta function interactions \cite{Olshanii,PetShlWal00}.
In the spatially uniform case 
the exact ground state was found in 1963 by Lieb and 
Liniger (LL) \cite{LL}. The properties of
this LL gas are determined by a dimensionless parameter 
$\gamma=mg_{1D}/n\hbar^2$ where $n$ is the 1D density and 
the effective 1D coupling constant
$g_{1D}$ is related to the 3D s-wave scattering length $a$ by \cite{Olshanii}
$g_{\rm 1D}
=2a\hbar{^2}/{m\ell_{0}^{2}
[1-({\mathcal C}/\sqrt{2})(a/\ell_{0})]}$
with $\ell_{0}=\sqrt{\hbar/m\omega_{0}}$ the transverse oscillator
length and ${\mathcal C}=1.4603\dots$. 
If $\gamma\gg 1$ the dynamics reduce to those of a 1D gas of hard
core, or impenetrable, point bosons \cite{Olshanii,PetShlWal00}, the 
``TG gas'' \cite{Olsh01,Ohberg,BMO02,OD02,GanShl02,AstGio02}.
This is a model for which exact many-body energy
eigensolutions and dynamics were found using a mapping from the
Hilbert space of eigenstates of an ideal gas of spinless
fermions to that of many-body eigenstates of hard core bosons 
\cite{map,map2,1dsho,soliton,breakdown,ring1,ring2,double-X}. 
The TG regime is only a tiny part of the LL regime, most of the latter 
having already been reached experimentally 
\cite{Bongs,Schreck,Greiner,DePMcCWin99,GorVogLea01}, 
and it is desirable to extend the 
techniques applicable in the TG regime to the whole LL regime
so as to reduce the \emph{many-body} dynamics of free and
trapped LL gases to solution of the time-dependent \emph{single-particle}
Schr\"{o}dinger equation for a set of time-dependent orbitals evolving
from initial orbitals comprising a LL-Fermi sea. It will be shown herein
that although this cannot be done exactly for all values of $\gamma$
and all time intervals, it can be done in good approximation for limited
time intervals for all $\gamma$, and exactly in some special
cases. 

{\it LL cusp condition and energy eigenstates:} The  
$N$-particle energy eigenstates satisfy 
$\hat{H}_{0}\psi_{B\alpha}=E_{\alpha}\psi_{B\alpha}$ 
where $\psi_{B\alpha}$ is symmetric (Bose). LL considered a uniform
system and 
used dimensionless units such that $\hbar=2m=1$, yielding a Hamiltonian
\begin{equation}\label{LL-Hamiltonian}
\hat{H}_{0}=\sum_{j=1}^{N}-\frac{\partial^2}{\partial x_{j}^2}
+2c\sum_{1\le j<k\le N}\delta(x_{j}-x_{k})
\end{equation}
where $2c=g_{1D}=2\gamma n$.
They used periodic boundary conditions
$\psi_{B\alpha}(\cdots x_{j}\pm L\cdots)=\psi_{B\alpha}(\cdots x_j\cdots)$.
The $N$-particle configuration space decomposes into $N!$
permutation sectors, each defined by one particular ordering
of numerical values of the $x_{1},\cdots,x_{N}$. In view of
the Bose symmetry of the $\psi_{B\alpha}$, it is sufficient to define them only
in the fundamental permutation sector $\mathfrak{R}_1$ wherein
$x_{1}<x_{2}<\cdots<x_{N}$, then extending to all $N!$
sectors by symmetry. The cusp condition following from
the $\delta(x_{j}-x_{k})$ interactions is \cite{LL} 
\begin{equation}\label{cusp}
\left(\frac{\partial}{\partial x_{j}}-\frac{\partial}{\partial x_{k}}\right)
\psi_{B\alpha}|_{x_{j}=x_{k}+}=c\psi_{B\alpha}|_{x_{j}=x_{k}+}\ .
\end{equation}
Define 
$\hat{B}_{j}=\partial/\partial x_{j+1}-\partial/\partial x_{j}-c\hat{1}$ 
where $\hat{1}$ is the unit operator. Then in $\mathfrak{R}_1$ 
\begin{equation}\label{B}
\hat{B}_{j}\psi_{B\alpha}|_{x_{j+1}=x_{j}+}=0 \ .
\end{equation}
In $\mathfrak{R}_1$ the energy eigenstates satisfying this are 
\begin{equation}\label{LL-eigenstates}
\psi_{B\alpha}(x_{1},\cdots,x_{N})=\frac{1}{\sqrt{N!L^{N}}}
\sum_{P}a_{P}(K_{\alpha})e^{i(PK_{\alpha})\cdot X}
\end{equation}
and their energies are $E_{\alpha}=\sum_{j}k_{j}^{2}$.
Here $K_{\alpha}$ is an N-dimensional vector $(k_{1},k_{2},\cdots,k_{N})$
with $k_{1}<k_{2}<\cdots<k_{N}$, the sum runs over the $N!$ permutations
$P$ of this set, and $X$ is the $N$-dimensional position vector
$X=(x_{1},x_{2},\cdots,x_{N})$, assuming $x_{1}<x_{2}<\cdots<x_{N}$.
There are infinitely many such choices $k_{1},\cdots,k_N$, and 
$\alpha$ labels each particular choice. 
The coefficients are \cite{GanShl02,Korepin}
\begin{equation}\label{a_P}
a_{P}(K_{\alpha})=\epsilon_{P}\prod_{1\le j<\ell\le N}
\frac{\gamma n+i(Pk_{\ell}-Pk_{j})}
{\sqrt{(\gamma n)^{2}+(Pk_{\ell}-Pk_{j})^{2}}}
\end{equation}
where $Pk_j$ is the image of $k_j$ under $P$ and
$\epsilon_P$ is the parity $\pm 1$ of $P$. This differs
from the $a_P$ of LL Eqs. (2.12) ff. only by a constant phase.

Although the symmetrical extension of $\psi_{B\alpha}$ to all permutation
sectors is bosonic (totally symmetric under permutations of 
$x_{1},\cdots,x_N$), it can be shown \cite{Korepin} that it 
is nevertheless \emph{anti}symmetric under 
permutations of $k_{1},\cdots,k_N$. Thus $\psi_{B\alpha}$ must be built from
$N$ \emph{singly-occupied} orbitals. 
The ground state $\psi_{B0}$ is obtained by a particular choice $K_0$ 
consisting of the smallest $N$ different $k_j$ consistent with the periodic 
boundary condition on the $\psi_{B\alpha}$ \cite{LL}, a ``LL-Fermi sea''.

{\it Thermodynamic limit:} The allowed $k_j$ were determined
in the LL paper by requiring that 
$\psi_{B\alpha}(x_{1},\cdots,x_{N})$ be periodic in each $x_j$
with periodicity length $L$, leading to 
$(k_{j+1}-k_{j})L=\sum_{s=1}^{N}(\theta_{sj}-\theta_{s,j+1})+2\pi n_{j}$
where $\theta_{sj}=-2\tan^{-1}[(k_{s}-k_{j})/\gamma n]$ 
and the $n_j$ are any integers $\ge 1$, whose
values distinguish between all the $N$-particle energy eigenstates 
\cite{LL,LL2}. The ground state is given by 
$n_{j}=1$ for all $j$. It follows that the individual orbitals $e^{ik_{j}x}$
in (\ref{LL-eigenstates}) are \emph{not} L-periodic; the 
$k_j$ are not actual particle momenta but rather nonuniformly spaced
quasimomenta \cite{LL}. Since our goal
is reduction of the many-particle time-dependent Schr\"{o}dinger equation
(MBTDSE) to the single-particle time-dependent Schr\"{o}dinger equation
(SPTDSE), this leads to difficulties related to nonhermiticity
of the single-particle Hamiltonian and an ill-posed initial value
problem for the orbital evolutes $\phi_{j}(x,t)$ on the finite interval
$-\frac{L}{2}<x<\frac{L}{2}$. These can be mitigated in the
thermodynamic limit $N\to\infty$, $L\to\infty$, $N/L\to n=\textrm{const.}$ 
wherein both the quasimomenta $k_j$ 
and the actual particle momenta $k_{\nu}$ become dense and every
quasimomentum can be approximated with arbitrary accuracy by the
$k_{\nu}=\nu 2\pi/L$ with $\nu=0,\pm 1,\pm 2,\cdots$, 
appropriate to periodic boundary conditions. The initial orbitals
$e^{ik_{\nu}x}/\sqrt{L}$ and their evolutes $\phi_{\nu}(x,t)$ are then
orthonormal on $-\frac{L}{2}<x<\frac{L}{2}$. In this limit
any thermodynamically intensive or extensive quantity expressible as
a sum of contributions of ``quasiorbitals'' evolving from
the quasimomentum plane waves $e^{ik_{j}x}$ approaches a similar
expression expressed in terms of contributions of L-periodic orbitals
evolving from L-periodic plane waves $e^{ik_{\nu}x}$. In doing
this it is necessary to take into account 
the nontrivial density of states of the allowed quasimomenta.
For example, the ground state energy in the
thermodynamic limit is 
$\sum_{\nu=-\nu_{0}}^{\nu_0}\beta(k_{\nu})k_{\nu}^{2}$ where 
$\beta(k_{\nu})$ is the ratio of the LL ground state density of states 
to its limiting value $1/2\pi$ in the TG limit, and $\nu_{0}=Lk_{LLF}/2\pi$ 
where $k_{LLF}$, the wave vector at the top of the LL-Fermi sea, is determined
by the condition $\int_{-k_{LLF}}^{k_{LLF}}\beta(k)dk=2\pi n$ \cite{LL}. 

{\it Response to phase imprinting:} Solitons have been generated
in experiments on cigar-shaped BECs by application of phase-imprinting pulses
\cite{Burger,Denschlag2}. We have recently studied \cite{soliton}
the exact dynamics of such a process in the TG limit, as well as 
phase-imprinting pulse generated interference effects
in a TG gas on a ring \cite{ring1,ring2} and in a double-X atom
interferometer \cite{double-X}. A similar LL model will be studied here. 
Consider a LL gas on a ring of circumference $L$ initially in its ground 
state, and represent the positions of the $N$ atoms by a
1D coordinate $-\frac{L}{2}<x<\frac{L}{2}$ measured around the circumference. 
Suppose that this system is subjected at $t=0$ to a phase-imprinting 
pulse $v(x,t)=-\delta(t)S(x)$. 
Denoting the orbitals of the LL solution 
by $e^{ik_{j}x}$, the orbitals just after the pulse are 
\cite{Rojo} $\phi_{j}(x,t=0+)=e^{iS(x)}e^{ik_{j}x}$, and by
(\ref{LL-eigenstates}) just after the pulse
$\psi_{B}(x_{1},\cdots,x_{N};t=0+)
=\psi_{B\alpha}(x_{1},\cdots,x_{N})\exp(i\sum_{j=1}^{N}S(x_{j}))$. Supposing
that $x_{1}<x_{2}<\cdots<x_{N}$, and using the $\hat{B}_j$ of 
Eqs. (\ref{cusp}) and (\ref{B}), one finds 
\begin{equation}\label{inv}
\hat{B}_{j}\psi_{B}(0+)=i\psi_{B}(0+)[S^{'}(x_{j+1})\!-\!S^{'}(x_{j})]
+e^{iS}\hat{B}_{j}\psi_{B\alpha}
\end{equation}
with $S=\sum_{j}S(x_{j})$. This 
vanishes as $x_{j+1}\to x_{j}+$. It follows that the cusp condition need not be
imposed as a constraint during the pulse, being exactly valid at $t=0+$ 
independently of the spatial form and strength of the pulse.

{\it Dynamics:} Consider next free propagation according to
$\hat{H}_{0}$ following such a pulse. This is generated 
in $\mathfrak{R}_1$ by the kinetic energy alone, i.e., by 
Eq. (\ref{LL-Hamiltonian}) with the delta
function interactions omitted. Let $\psi_{B}(x_{1},\cdots,x_{N};t)$ be 
a wave function differing from (\ref{LL-eigenstates}) only by replacement 
of the orbitals $e^{ik_{j}x}$ by the solutions $\phi_{j}(x,t)$
of the \emph{single}-particle TDSE 
$[i\partial/\partial t +\partial^{2}/\partial x^{2}]\phi_{j}(x,t)=0$
reducing to $\phi_{j}(x,t=0+)$ as $t\to 0+$. 
Then $\psi_{B}(x_{1},\cdots,x_{N};t)$ satisfies the 
MBTDSE $[i\hbar\partial/\partial t-\hat{H_0}(t)]\psi_{B}=0$ in the
\emph{interior} of $\mathfrak{R}_1$ for $t>0$ and reduces to 
the previously discussed wave function 
$\psi_{B}(x_{1},\cdots,x_{N};t=0+)$ as $t\to 0+$. 
However, a subtle complication enters here: 
Although the derivative operators in $\hat{B}_j$ of Eq. (\ref{cusp})  
commute with the
kinetic energy, the ratio of third to
second derivative of the wave function is different from 
the ratio of first to zero{\it th} derivative required by the cusp
condition \cite{LL,OD02}, so a wave function propagated as above 
does not in general 
satisfy the cusp condition for $t>0$. However, it is satisfied trivially for 
$\gamma\to 0$, and also for $\gamma\to\infty$ (TG limit). Furthermore, by
temporal continuity of the solution of the TDSE it is satisfied in good
approximation for a nonzero range of times $t>0$ for all $\gamma$, 
assuming as in (\ref{inv}) that $\psi_{B}(\cdots;t=0)$ is an energy 
eigenstate (\ref{LL-eigenstates}), (\ref{a_P}). By a temporal phase mixing
argument the length of this range should increase with decreasing values
of $\Delta E/E$, suggesting that constraint-free propagation as above
is a reasonable approximation for $\Delta E/E\ll 1$ where $E$ is the energy and
$\Delta E$ is its dispersion. 

{\it Single particle density and soliton propagation:} Suppose now that
the initial $N$-particle wave function $\psi_{B}(\cdots;t=0+)$
differs from the LL ground state $\psi_{B0}$ at most by multiplication of the
LL orbitals $e^{ik_{j}x}/\sqrt{L}$ by $e^{iS(x)}$ as in the previously
discussed case of phase imprinting. Then $\psi_{B}(\cdots;t=0+)$ will
satisfy the LL cusp condition exactly. Suppose furthermore that 
$\Delta E/E\ll1$ where $\Delta E$ is the energy
uncertainty of $\psi_{B}(\cdots;t=0+)$. Then to a good approximation 
one has in $\mathfrak{R}_1$
\begin{eqnarray}\label{evolution}
&&\psi_{B}(x_{1},\cdots,x_{N};t)\nonumber\\
&&=\frac{1}{\sqrt{N!}}
\sum_{P}a_{P}(K_{0})\phi_{P1}(x_{1},t)\cdots \phi_{PN}(x_{N},t)
\end{eqnarray}
where the $\phi_{j}(x,t)$ are evolved by the single-particle TDSE starting 
from $\phi_{j}(x,0+)=e^{ik_{j}x}e^{iS(x)}/\sqrt{L}$. 
The single particle density at time $t$ is
$n(x,t)
=N\int\cdots\int|\psi_{B}(x,x_{2},\cdots,x_{N};t)|^{2}dx_{2}\cdots dx_N$
where the integrals run over one periodicity cell $-L/2\le x_{j}\le L/2$ 
and the thermodynamic limit is assumed so that periodic exponential orbitals 
are used.  Upon substitution from (\ref{evolution})
one obtains a double permutation sum $\sum_{PP'}$. Consider first the
terms $P\ne P'$. Recall that (\ref{evolution}) holds
only in $\mathfrak{R}_1$ and is to be extended
to other sectors by Bose symmetry. Then since $P'$ differs from $P$ by
at least one interchange, and in most cases by many, integrals of the
form $\int_{-L/2}^{x_p}\phi_{j}^{*}(x_{m},t)\phi_{\ell}(x_{m},t)dx_m$
and $\int_{x_p}^{L/2}\phi_{j}^{*}(x_{m},t)\phi_{\ell}(x_{m},t)dx_m$
arise, where $j$, $l$, $m$, and $p$ are all different. The phases of
these integrals have denominators $L$ following from the normalization of
the $\phi_j$ and their phases oscillate as a function of $j-\ell$, and this
phase variation is compounded by oscillation of the phases of the
factors $a_{P}^{*}(K_{0})a_{P'}(K_{0})$ from Eq. (\ref{a_P}). One therefore
expects that in the thermodynamic limit the net off-diagonal contribution
will be small compared with the diagonal ones and will be dropped,
an approximation in the same spirit as the random phase
approximation. The diagonal contributions $P=P'$ behave very
differently. Since (\ref{evolution}) holds only in $\mathfrak{R}_1$ 
and is to be extended to other permutation sectors by Bose symmetry,
the integrations over $x_{2},\cdots,x_N$ have different limits for each
different ordering of $x,x_{2},\cdots,x_N$. However, summing over all
permutations, collecting coefficients of $|\phi_{1}(x)|^{2}$, and 
performing appropriate permutations of the names of integration variables, 
one finds that the integrations over $x_{2},\cdots,x_N$ sum to a
multiple of 
$\int_{-L/2}^{L/2}dx_{2}\cdots\int_{-L/2}^{L/2}dx_{N}
|\phi_{2}(x_{2})|^{2}\cdots|\phi_{N}(x_{N})|^{2}=1$, and the same holds
for the coefficients of the other $|\phi_{j}(x)|^{2}$. Noting also that 
$|a_{P}(K_{0})|^{2}=1$ by (\ref{a_P}) and that there are $(N-1)!$
permutations $P$ such that $P1=j$ for each $j=1,\cdots,N$, one finds that
the final expression collapses simply to the sum of
absolute squares of all $N$ orbitals. Recalling that approximation of
the nonperiodic LL exponentials $e^{ik_{j}x}$ by the periodic exponentials
$e^{ik_{\nu}x}$ in the thermodynamic limit requires inclusion of a density
of states factor, one obtains
\begin{equation}\label{density}
n(x,t)=\sum_{\nu=-\nu_{0}}^{\nu_0}\beta(k_{\nu})|\phi_{\nu}(x,t)|^2 
\end{equation}
where $\nu_0$ and $\beta(k_{\nu})$ were defined in the previous discussion
of the thermodynamic limit.

\begin{figure}
\includegraphics[width=1.0\columnwidth,angle=0]{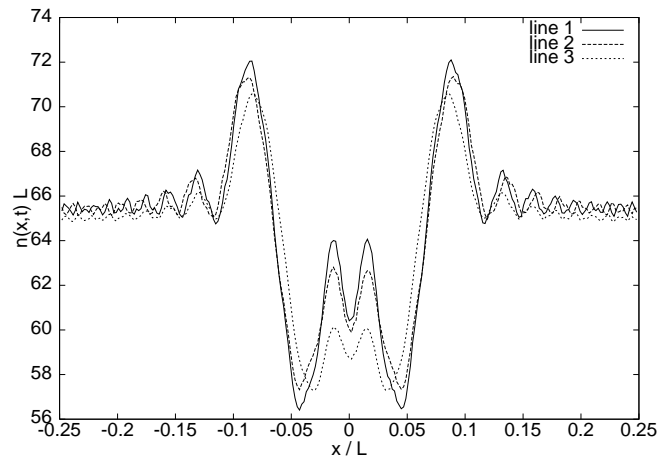}
\caption{Scaled density $n(x,t)L$ versus scaled position $x/L$ around the
ring using $65$ orbitals, at time $t$ after a $\pi$ phase imprint
imposed for $-L/2\le x\le 0$. The middle half $-L/4\le x\le L/4$ of
the periodicity cell is shown. Line 1: $\gamma=1000$ (TG limit), $L=1$,
$t=.001\tau$. Line 2: $\gamma=1$, $L=2.19$, $t=.005\tau$. Line 3: $\gamma=.05$
(Bogoliubov regime), $L=7.73$, $t=.06\tau$. Here $\tau=1/2\pi$ in the units
used ($\hbar=2m=1$, mean density $n=65$).}
\label{Fig:one}
\vspace{-0.5cm}
\end{figure}

To apply this to gray soliton generation consider a ring of circumference $L$ 
with circumferential coordinate $x$ with $-L/2\le x\le L/2$ 
and impose a phase shift $\pi$ on the left half $-L/2\le x\le 0$.
The density of states was determined by 
solution of Eqs. (3.18) and (3.20) of LL \cite{LL,Atkinson}, and the
TDSE propagation was effected by FFT of the initial orbitals
$e^{ik_{\nu}x}e^{iS(x)}$ after which the resultant Fourier components have 
trivial time dependence $e^{-ik^{2}t}$. The density functions showing 
soliton propagation are shown in Fig. \ref{Fig:one} for the cases 
$\gamma=\infty$ (TG limit), $\gamma=1$ (intermediate regime), and 
$\gamma=0.05$ (Bogoliubov regime) for $\nu_{0}=32$, corresponding to a 
LL-Fermi sea of $65$ orbitals. The periodicity cell lengths
$L$ were adjusted to keep the mean density ($65$ in the units used)
and hence the LL-Fermi momentum fixed, using Eqs. (3.14)-(3.21) of
LL \cite{LL}. The shape of the density profile is 
quite insensitive to $\gamma$ apart from 
the indicated scaling of the periodicity length $L$ and observation 
time $t$, in spite of strong peaking of the density of states 
about low momenta as $\gamma$ varies from the TG regime $\gamma\gg 1$
to the Bogoliubov regime $\gamma\ll 1$. 
This insensitivity is consistent with scaling of the allowed momenta
$k_{\nu}=2\pi/L$ inversely with $L$ and scaling of observation times $t$ 
in their propagation factors $e^{-ik_{\nu}^{2}t}$ directly with $L^2$.

\begin{figure}
\includegraphics[width=1.0\columnwidth,angle=0]{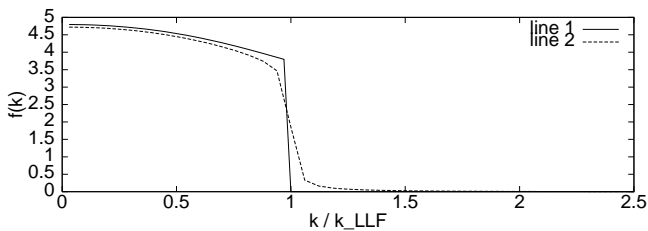}
\caption{Quasimomentum distributions $f(k)$ for $\gamma=1$,
normalized to total number of particles nL, with n and L
the same as in Fig. 1.   
Line 1: Unperturbed LL-Fermi sea. Line 2: Perturbed sea after a 
$\pi$ phase imprint applied to half of the ring.}
\label{Fig:two}
\vspace{-0.5cm}
\end{figure}

One can define a LL-Fermi sea quasimomentum distribution
function $f(k)$ as the absolute square of the Fourier transform of
each orbital, weighted by the LL quasimomentum density of states and
summed over all orbitals. This was evaluated with the aid of an FFT code and 
is shown in Fig. \ref{Fig:two} for the
case $\gamma=1$ before and just after the $\pi$ phase imprint.The 
perturbation promotes particles just
under the LL-Fermi surface to states just above the surface, indicating
that Pauli blocking is still quite effective at $\gamma=1$, to be
expected since the density of states is still almost flat. The
quasimomentum distribution is not the same as the true momentum distribution,
which is peaked about $k=0$ even in the TG limit \cite{Olshanii} and becomes 
increasingly peaked as $\gamma$ decreases \cite{AstGio02}. Nevertheless,
the energy is given exactly by $E=\sum_k k^{2}f(k)$ both for the LL ground
state \cite{LL} and the LL excited states \cite{LL2}. It is thus 
reasonable to approximate the energy of the phase-imprinted state by the same 
simple expression. Furthermore, since one starts from the ground state which 
has zero dispersion, the energy dispersion after the pulse should be of
the same order as the mean excitation energy 
$E_{after}-E_{before}$. The FFT evaluation yields 
$(E_{after}-E_{before})/E_{before}=0.083\ll 1$, suggesting that the soliton 
density profiles of Fig. \ref{Fig:one} are good approximations to exact
profiles which are not currently calculable. 

\emph{Discussion:} The LL solitons exhibit density
ripples absent from Gross-Pitaevskii solitons even for $\gamma$ as small 
as $.05$, being many-orbital features stemming from fermionization 
which is present for any nonzero $\gamma$, no matter how small 
\cite{LL,Korepin}. Even the ground state of a trapped LL gas is currently
unknown, so it is not clear whether or not such ripples will be present
in soliton-like dynamics of a trapped LL-gas, an important topic for future
investigations. 
The approach used for $\delta(t)$ pulses could be extended
to arbitrary potentials $v(x,t)$ by split-operator propagation of orbitals.
Could it be extended to the trapped LL gas? 
\begin{acknowledgments}
I am grateful to Maxim Olshanii and Dimitri Gangardt for helpful comments.
I thank Ewan Wright for such
comments and for copies of his FFT TDSE solver and quasimomentum
distribution codes. This work was begun at the 
Institute for Theoretical Atomic and Molecular Physics, Harvard-Smithsonian 
Center for Astrophysics, 
where it was partially supported by the National Science Foundation,
and at the Benasque, Spain 2002 workshop Physics of Ultracold Dilute 
Atomic Gases. I thank Kate Kirby for the hospitality of ITAMP and 
Anthony Leggett and Fernando Sols for that of the Benasque workshop.
This work has been  supported at the 
University of Arizona by Office of Naval Research grant N00014-99-1-0806.
\end{acknowledgments}
\end{document}